\documentclass[11pt,twoside]{article}
\usepackage{asp2014}

\aspSuppressVolSlug
\resetcounters

\begin{document}

\title{Spacecraft Telecommunications}
\author{T.~Joseph~W.~Lazio$^1$ and Sami Asmar${}^1$
\affil{$^1$Jet Propulsion Laboratory, California Institute of
  Technology, Pasadena, \hbox{CA}, \hbox{USA}; \email{Joseph.Lazio@jpl.nasa.gov}}
}

\paperauthor{T.~Joseph~W.~Lazio}{Joseph.Lazio@jpl.nasa.gov}{}%
            {Jet Propulsion Laboratory, California Institute of
              Technology}%
            {Interplanetary Network
              Directorate}{Pasadena}{CA}{91109}{USA}
\paperauthor{Sami Asmar}{Sami.Asmar@jpl.nasa.gov}{}%
            {Jet Propulsion Laboratory, California Institute of
              Technology}%
            {Interplanetary Network Directorate}{Pasadena}{CA}{91109}{USA}

\begin{abstract}
There is a long history of radio telescopes being used to augment the
radio antennas regularly used to conduct telemetry, tracking, and
command of deep space spacecraft.  Radio telescopes are particularly
valuable during short-duration mission critical events, such as
planetary landings, or when a mission lifetime itself is short, such
as a probe into a giant planet's atmosphere.  By virtue of its high
sensitivity and frequency coverage, the next-generation Very Large
Array would be a powerful addition to regular spacecraft ground
systems.  Further, the science focus of many of these deep-space
missions provides a ``ground truth'' in the solar system that
complements other aspects of the ngVLA's science case, such as the
formation of planets in proto-planetary disks.
\end{abstract}

\section{Introduction}\label{lazio.sec:intro}

In addition to the multitude of astronomical applications described
elsewhere in this volume, the next generation Very Large Array (ngVLA)
could provide valuable benefits in other fields such as the
telecommunications with robotic spacecraft on scientific missions
throughout the Solar System.  There have been many cases of radio
astronomical telescopes participating in and enabling planetary
missions by augmenting the receiving capabilities of traditional
spacecraft tracking facilities, with notable examples including the VLA's
participation in Voyager~2's flyby of the planet Neptune \citep[Figure~\ref{fig:lazio.neptune},][]{lb85}
and the use of the Green Bank Telescope to track the descent of the
Huygens probe to the surface of Titan and enabling the Doppler Wind
Experiment \citep{fab+06}.
The sensitivity and frequency coverage of the ngVLA has the potential
to exceed considerably that of any existing or planned dedicated
spacecraft radio tracking facility.  This contribution is focused on
the relevance of the ngVLA for potential NASA missions, but, with the increasing
number of space agencies conducting deep space missions,\footnote{
In addition to \hbox{NASA}, the European Space Agency (ESA), the
Japanese Aerospace Exploration Agency (JAXA), the Indian Space
Research Organization (ISRO), and the Roscosmos State Corporation for
Space Activities (Roscosmos) send robotic spacecraft into deep space,
and potentially Korean and United Arab Emirates will do so in the
future.}
the ngVLA could be of benefit to many space agencies for increasing
the science return from their space missions.

\articlefigure{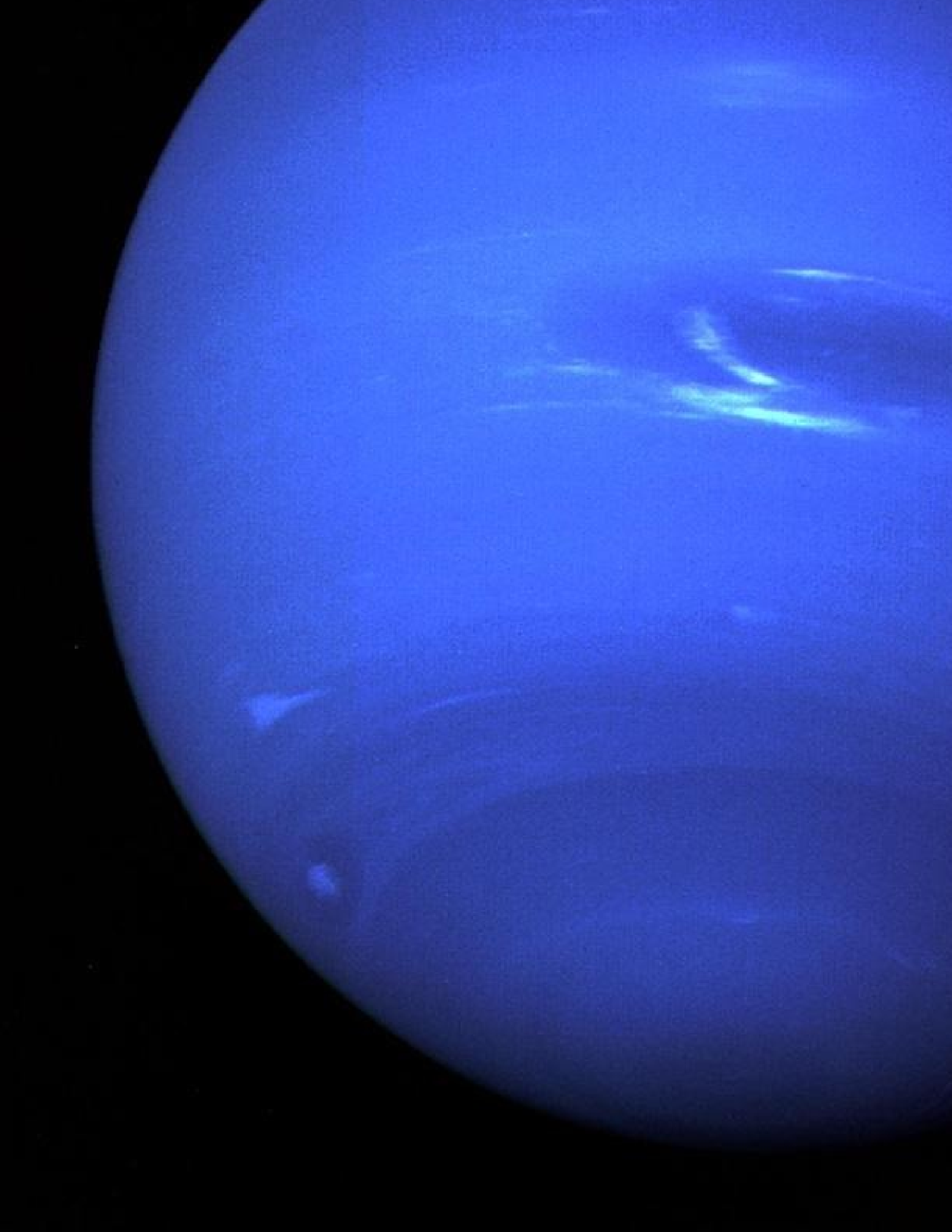}{fig:lazio.neptune}{%
  Neptune as viewed by Voyager~2 during its encounter in~1989.  This
  image is a combination of images acquired with the green and orange
  filters from the narrow angle camera, acquired on~1989 August~20,
  just under five days from closest approach.  Clearly visible is the
  Great Dark Spot.  These images were likely downloaded via a
  combination of telemetry downlinks from the Very Large Array (VLA)
  and NASA's Deep Space Network.  In a similar manner, the ngVLA could
  enable future mission critical events by robotic spacecraft
  throughout the solar system.}

\section{Scientific Context}\label{lazio.sec:context}

Historically, NASA's most demanding application for highly sensitive
ground receiving stations has been planetary science missions, in
which spacecraft have been sent to other bodies in the solar system,
often at distances of multiple astronomical units from the Earth.
Increasingly, however, heliophysics and astrophysics
missions have begun to demand similarly sensitive ground receiving
stations.  For instance, the Solar TErrestrial RElations
Observatory (STEREO) mission has two spacecraft, one in an
Earth-leading orbit and one in an Earth-trailing orbit, that are
approximately 1~au distant\footnote{%
  At the time of writing, the STEREO-B spacecraft is not operational.}
and the Parker Solar Probe\footnote{%
  At the time of writing, scheduled for launch in~2018 August.}
will dip to within a few solar radii of the Sun (i.e., travel to
approximately 1~au from the Earth).  In order to reduce
interference from the Earth and operate in a stable environment, many
astrophysics missions (e.g., Planck, \textit{Spitzer} Space Telescope,
\textit{Kepler},
\textit{James Webb Space Telescope}) are being launched into orbits
such as Earth-trailing orbits or Earth-Sun Lagrange~2 point halo
orbits.

In this landscape, the ngVLA would complement NASA and ESA's network
of deep space antennas by providing additional sensitivity for
short-duration, high priority events.  NASA's Deep Space Network (DSN)
is a network of~34~m and~70~m diameter antennas at three longitudes
around the world that provides routine collection of telemetry data
(a.{}k.{}a.\ ``downlink'') from spacecraft beyond geosynchronous
orbit; ESA has three 35~m antennas at similar longitudes with similar
functionality.  However, it is standard practice to consider
augmenting these antennas during mission critical events, such as the
entry, descent, and landing of a spacecraft on another planet.  Such
augmentation provides resiliency for the mission; for instance, for
the Doppler wind experiment on the Huygens probe to Titan, the
receiver on-board the \textit{Cassini} spacecraft was not configured
correctly and only the data from Earth-based radio telescopes assured
the success of that experiment \citep{fab+06}.  In addition, with its
higher sensitivity, the ngVLA may enable the collection of additional
data.

Table~\ref{tab:sensitivity} compares the relative performance of the
current DSN antennas (DSN~810-005 Modules~101, 104) with that
potentially achievable with a fraction of the \hbox{ngVLA}, using the
antenna gain at two standard deep space communications bands, X~band
(receiving at~8.24~GHz) and Ka~band (receiving at~32~GHz); a 34~m
antenna operating at X band is taken to be the reference.  For the
\hbox{ngVLA}, only the Core Array \citep{c18}, consisting of~94
antennas of~18~m diameter, is considered; the antennas are assumed to
have an efficiency of~65\%.  While the actual performance of the ngVLA
may differ from that shown in Table~\ref{tab:sensitivity}, it is clear
that even a fraction of the full ngVLA offers substantial
improvements.  All other things being equal, the relative performance
of Table~\ref{tab:sensitivity} is equivalent to the signal-to-noise
ratio achieved on the downlink signal, as the DSN antennas have system
temperatures of approximately 20~K and we assume similar values for
the ngVLA antennas.  In turn, signal-to-noise ratio translates
approximately linearly to downlink data rate; alternately, the higher
sensitivity can be considered additional margin against unexpected
events.  We now discuss a few examples of mission concepts that could
benefit from having extremely high sensitivity ground receiving
stations.

\begin{table}[bth]
  \begin{center}
    \caption{DSN and ngVLA Receiving Antenna Relative Performance}
  \label{tab:sensitivity}
  \begin{tabular}{lcc}

    \tableline
    \noalign{\smallskip}
    \textbf{System} & \multicolumn{2}{c}{\textbf{Relative Performance}} \\
                    & \textbf{X~band} & \textbf{Ka~band} \\
    \noalign{\smallskip}
    \tableline
    \noalign{\smallskip}
    DSN 34~m                 & 1.0        & 11.7 \\
    DSN 70~m                 & 4.3        & \ldots \\
    DSN 4 $\times$ 34~m      & 3.7        & 46.8 \\
    ngVLA Core Array         & 17.1       & 248 \\
    (94 antennas, 18~m diameter) \\
    \noalign{\smallskip}
    \tableline
  \end{tabular}
  \end{center}
\end{table}

The \textit{Visions and Voyages for Planetary Science in the Decade
  2013--2022} Decadal Survey describes two New 
Frontier\footnote{%
  NASA's medium-class missions, with a cost cap of approximately \$1~billion.}
mission concepts that would necessarily be short-lived.  The Saturn
Probe concept would deploy a probe into Saturn's atmosphere to measure
its structure and composition.  While the materials used to construct
such a probe may have improved since the construction of the Galileo
probe \citep[and references within]{yss96}, the Saturn probe would suffer the same eventual fate, crushed
under the rising pressures as it descends.  The Venus In Situ Explorer
concept would land on the surface of Venus in order to sample the
composition of the crust.  Again, while material properties and
thermal control systems have improved since the construction Soviet
Venera landers, the surface of Venus is such a harsh environment that
a lander will have a necessarily finite lifetime.

The \textit{Ice Giants Pre-Decadal Survey Mission Study Report}
considers flagship-class mission concepts to Uranus, Neptune, or both.
Four mission architectures were studied in detail, with three of the
four including an atmospheric probe.  (Two architectures involved an
atmospheric probe for Uranus, one for Neptune.)  The expected lifetime
of such a probe in the atmosphere of either planet is approximately
1~hr.

During the last perihelion passage of Comet~1/P Halley, a figurative
armada of spacecraft were sent to conduct close fly-bys of its
nucleus.  Because of the high relative velocities between the Earth
and a long-period comet such as Halley, the durations of the fly-bys
were short.  Inspired by both the success of the spacecraft that flew
by Halley and the recent discovery of the first interstellar object
\citep[1I/2017~U1 `Oumuamua,][]{mwm+17}, there is active consideration of what kind of
mission concept or concepts might be possible to enable a similar
encounter.  While the relative velocities could be even higher
than in the case of the Halley encounters, the potential science
return from a close encounter with an interstellar object would likely
be significant.

These mission concepts are intended to be illustrative only, and
specific instruments have not been selected.  Nonetheless, the higher
data rates enabled by the use of the ngVLA could enable more capable
instruments (e.g., higher resolution mass spectrometers).  Further,
these example mission concepts are certainly not an exhaustive list of
all possible short-lived mission concepts, but they illustrate how
high-priority missions might nonetheless be short-lived and benefit
greatly from additional sensitivity such as could be provided by the
\hbox{ngVLA}.

An additional benefit of the ngVLA may be for use during spacecraft
emergencies.  It is standard practice to include low-gain antennas on
deep space spacecraft for the purposes of enabling communications even
if the spacecraft's attitude is unknown or uncontrolled.  Low-gain
antennas have the obvious benefit of a wide field of view at the cost
of sensitivity.  The received signal from a spacecraft in ``safe
mode'' would most likely be (much) weaker than planned from its
high-gain antenna, for which the ngVLA's high sensitity could be of
benefit in capturing engineering data and understanding the problem
with the spacecraft.  

\section{Design Considerations}\label{lazio.sec:design}

The primary requirement for the ngVLA to receive spacecraft telemetry
is for it to have frequency coverage of the  allocated space-to-Earth
downlink bands (Table~\ref{tab:ttc.bands}).  The relevance of the
ngVLA is immediately obvious, as its planned frequency coverage
encompasses all of the frequency allocations.

\begin{table}[tbh]
\centering
\caption{Frequency Bands for Deep-Space Spacecraft Telemetry\label{tab:ttc.bands}}
\smallskip
\begin{tabular}{lc}
\tableline
\noalign{\smallskip}
\textbf{Name} & \textbf{Telemetry (Downlink)} \\
             & (space-Earth) \\
             & (GHz)         \\
\noalign{\smallskip}
\tableline
\noalign{\smallskip}
S~band       & 2.29--2.30 \\
X~band       & 8.40--8.45 \\
Ka~band      & 31.8--32.3 \\
\noalign{\smallskip}
\tableline
\end{tabular}
\end{table}

Two additional bands warrant comment.  There is a space-to-Earth
allocation in the K~band (25.5~GHz--27.0~GHz).  This band is allocated
for spacecraft with orbits that take them no farther than $2 \times
10^6$~km from Earth.  This distance includes the Earth-Sun~L2
point, which is used increasingly by astrophysics missions such as the
\textit{James Webb Space Telescope}.  While such missions would also
benefit from high sensitivity, they have regular telemetry
requirements, which are more likely to be met by NASA's DSN (or ESA's
Deep Space Antennas).  However, should such a spacecraft enter ``safe
mode,'' the ngVLA could assist with the recovery of the spacecraft.

There is also an allocation in the ultra-high frequency (UHF) portion
of the radio spectrum used for spacecraft-to-spacecraft communications
(``proximity links'').  This band is not within the ngVLA's frequency
coverage, and the SKA1-Mid or a similar facility would be more appropriate.

Like pulsars, spacecraft are point sources, and full field
interferometric synthesis is not required.  Beamforming suffices.  In
general, the requirements for pulsar beamformed observations would be
sufficient for spacecraft telemetry reception, with the added benefit
that spacecraft signals are typically quite narrow band relative to
astronomical sources.  A specific concern with the legacy VLA for both
pulsar observations and spacecraft telemetry was the ``waveguide
switch cycle,'' which resulted in a 1.5~ms data gap every 50~ms
\citep{d82,h99}; modern digital electronics and fiber optic
transmission should obviate this particular concern.

\section{Conclusion}\label{lazio.sec:conclude}

In conclusion, the ngVLA would be well suited for occasional use of
high priority space missions, particularly those with intrinsically
short lifetimes.  Not only can these missions be considerable
investments by a space agencies (with costs in excess of~\$1~billion),
the science motivations for them often complement the larger ngVLA
science case.  Most notably, the planets in our solar system form
the end states of the proto-planetary disks that the ngVLA will
image and only by studying both will we understand planet formation
and evolution.

\acknowledgements We thank D.~Jones, L.~Deutsch, and R.~Preston for helpful
discussions and guidance on this topic.  Some of the information
presented is pre-decisional and is intended for informational purposes
only.  Part of this work was carried out at the Jet Propulsion
Laboratory, California Institute of Technology, under contract with
the National Aeronautics and Space Administration.

\end{document}